# mcLARO: Multi-Contrast Learned Acquisition and Reconstruction Optimization for simultaneous quantitative multi-parametric mapping

Jinwei Zhang[1,2], Thanh D. Nguyen[2], Eddy Solomon[2], Chao Li[2,4], Qihao Zhang[2], Jiahao Li[1,2], Hang Zhang[3], Pascal Spincemaille[2], Yi Wang[1,2]*

[1]Department of Biomedical Engineering, Cornell University, Ithaca, NY, USA

[2]Department of Radiology, Weill Cornell Medicine, New York, NY, USA

[3]Department of Electrical and Computer Engineering, Cornell University, Ithaca, NY, USA

[4]Department of Applied Physics, Cornell University, Ithaca, NY, USA

* Correspondence to:

Yi Wang, PhD.

Radiology, Weill Cornell Medicine

407 E 61st St, New York, NY 10065, USA.

E-mail: yiwang@med.cornell.edu

# ABSTRACT

**Purpose:** To develop a method for rapid sub-millimeter T1, T2, T2* and QSM mapping in a single scan using multi-contrast Learned Acquisition and Reconstruction Optimization (mcLARO).

**Methods:** A pulse sequence was developed by interleaving inversion recovery and T2 magnetization preparations and single-echo and multi-echo gradient echo acquisitions, which sensitized k-space data to T1, T2, T2* and magnetic susceptibility. The proposed mcLARO used a deep learning framework to optimize both the multi-contrast k-space under-sampling pattern and the image reconstruction based on image feature fusion. The proposed mcLARO method with $R = 8$ under-sampling was validated in a retrospective ablation study using fully sampled data as reference and evaluated in a prospective study using separately acquired conventionally sampled quantitative maps as reference standard.

**Results:** The retrospective ablation study showed improved image sharpness of mcLARO compared to the baseline network without multi-contrast sampling pattern optimization or image feature fusion, and negligible bias and narrow 95% limits of agreement on regional T1, T2, T2* and QSM values were obtained by the under-sampled reconstructions compared to the fully sampled reconstruction. The prospective study showed small or negligible bias and narrow 95% limits of agreement on regional T1, T2, T2* and QSM values by mcLARO (5:39 mins) compared to reference scans (40:03 mins in total).

**Conclusion:** mcLARO enabled fast sub-millimeter T1, T2, T2* and QSM mapping in a single scan.

# INTRODUCTION

For developing MRI pulse sequences and image reconstructions in fast quantitative multi-parametric mapping, in addition to T1 and T2 relaxation time, there has been an increasing interest in incorporating multi-echo gradient echo (mGRE) acquisition into multi-contrast sequences to allow $T_2^*$ and quantitative susceptibility mapping (QSM) (1-3). QSM (4) is a post-processing technique which estimates the tissue local field from the total field derived from the GRE phase data (5,6) by applying background field removal (7) and performs the dipole field inversion to calculate the tissue susceptibility map (8). In the brain, QSM can provide a quantitative measure of local susceptibility sources in both healthy and pathological tissues including endogenous deoxyheme and ferritin iron, myelin, and calcium, as well as exogenous gadolinium- or iron-based contrast agents (9-11). QSM has been applied in multiple sclerosis (9), stroke and small vessel disease (12), and neurodegenerative disorders including amyotrophic lateral sclerosis(13), Parkinson's disease (14,15) and Alzheimer's disease (16).

Recent works incorporating T2* and QSM into multi-parametric mapping include MP2RAGEME (1), Multitasking (2) EPTI (3) and BUDA-SAGE (17). A common acquisition strategy of these methods is to combine magnetization preparations (e.g., inversion recovery (IR) for T1 weighting and T2prep for T2 weighting) with an mGRE readout using k-space under-sampling. Images are reconstructed using parallel imaging methods such as SENSE (18) or GRAPPA (19), or low-rank denoising (20) and multitasking (21). While promising, these methods suffer from relatively long scan time (1,2), low SNR on QSM due to the short last echo time (2), and thick slices with limited coverage (3), which limits their clinical utility.

Recently, deep learning approaches have been applied to k-space under-sampling pattern optimization (22,23), image reconstruction (24-30) and biophysical inverse problems (31-37) in MRI. For the under-sampling pattern optimization, LOUPE (22) and its extension LOUPE-ST (23) learned an optimal variable density k-space under-sampling pattern through back propagation, where a probabilistic density function of k-space data was updated to improve reconstructed image quality. For image reconstruction, MoDL (24) and VarNet (25) incorporated the parallel imaging forward model into the unrolled reconstruction networks, where convolutional denoisers were learned from fully sampled images to help reduce noise and aliasing artifacts from under-sampled reconstruction. In addition to the single-echo k-space imaging involved in the above methods, k-t imaging with multiple echoes, contrasts or frames has been accelerated using deep learning as well (28-30,38).

In this work, we extend our prior work, learned acquisition and reconstruction optimization (LARO) (28,30) for QSM acceleration, to multi-parametric mapping acceleration by 1) developing an IR and T2-prepared single and multi-echo GRE sequence for simultaneous T1w, T2w and T2*w signal acquisition, 2) building a multi-contrast under-sampling pattern and unrolled image reconstruction network optimization for accelerated imaging, and 3) deriving T1 and T2 maps using dictionary matching and T2* and QSM maps using multi-echo GRE signal fitting. The resulting method is named mcLARO: Multi-Contrast Learned Acquisition and Reconstruction Optimization.

## METHODS

### Pulse sequence design

Figure 1a shows the proposed multi-contrast pulse sequence inspired by the MP2RAGEME (1) and 3D-QALAS (39,40) designs. A module consisting of a non-selective inversion pulse followed by single-echo ($N_{GRE}$ TRs) and multi-echo ($N_{mGRE}$ TRs, each $N_E$ echoes) GRE readouts was used to sensitize T1, T2$^*$ and magnetic susceptibility. A second module consisting of a T2prep pulse (41,42) followed by a single-echo GRE readout was used for T2 relaxation measurement. Acquisition parameters for the proposed sequence are listed in Table S1.

### K-space sampling

Similar to the prospective under-sampling strategy in LARO (30), a radial fan-beam sampling scheme (43) is used for both fully sampled and variable density under-sampled scans in the proposed sequence. During the fully sampled scan with $258 \times 160$ $k_y - k_z$ acquisition matrix, k-space locations are divided into 244 fan-beam segments with 128 TRs in each segment for one repetition of the sequence, yielding $(244 * 128)/(258 * 160) = 75.7\%$ elliptical k-space coverage to acquire fully sampled data. At each repetition, in-and-out (k-space center in the middle of the segment), reverse-centric (k-space center at the end of the segment), and centric (k-space center at the beginning of the segment) ordering strategies are applied to the inversion recovery (IR) prepped single-echo, multi-echo, and T2 prepped single-echo acquisitions, respectively. The total scan time of the fully sampled data is 34:30 mins. The under-sampling pattern follows the same fan-beam strategy with sparsely sampled k-space locations. An $R = 8$ under-sampling pattern was implemented into the proposed sequence with 40 repetitions, resulting in $(40 * 128)/(258 * 160) = 12.40\%$ sampling ratio with a total scan time of 5:39 mins. More details regarding the fan-beam sampling strategy can be found in Figure 2 in (30).

**Optimized multi-contrast sampling pattern**

A sampling pattern optimization module (Figure 2b) proposed in (30) is used to optimize an $R = 8$ under-sampling pattern for each echo from the fully sampled k-space data. This module updates the variable density of the under-sampling pattern by learning a probabilistic density distribution that the under-sampling pattern is generated from. Specifically, 2D variable density Cartesian sampling patterns in the $k_y - k_z$ plane as shown in Figure 2b are used to retrospectively under-sample the fully sampled k-space data during the sampling pattern optimization process, where for the $j$-th contrast ($j = 1, 2, ..., N_E + 3$), learnable weights $w_j$ generate a probabilistic density pattern $P_j$ through sigmoid transformation with a slope parameter $a = 0.25$:

$$P_j(w_j) = \frac{1}{1 + e^{-a \cdot w_j}}. \tag{1}$$

Then a binary under-sampling pattern $U_j$ is generated via stochastic sampling from $P_j$ with indicator function $\mathbf{1}_x$ and sample $z$ from uniform distribution on $[0, 1]$:

$$U_j(w_j) = \mathbf{1}_{z < P_j(w_j)}. \tag{2}$$

A straight-through estimator (44) is used to overcome the zero gradient problem when backpropagating through Eq. 2:

$$\frac{d\mathbf{1}_{z < P_j(w_j)}}{dw_j} \rightarrow \frac{dP_j(w_j)}{dw_j}. \tag{3}$$

**Optimized multi-contrast reconstruction**

A deep ADMM network (Figure 2a) proposed in (30) is used for image reconstruction by unrolling an ADMM iterative scheme of multi-contrast images, where single and multi-echo images are reconstructed together. A multi-contrast feature fusion module (Figure 2c) is proposed by extending the temporal feature fusion module in (30) to aggregate features across single-echo and multi-echo contrasts during reconstruction. First, the temporal feature fusion module proposed in (30) is used to extract the features of the multi-echo images $s_j$ ($j = 1, 2, ..., N_E$) for T2* and QSM, where a recurrent convolutional network including convolutional layers $N_m(\cdot)$ and $N_h(\cdot)$ for $s_j$ and $h_j$ is used to generate the $j$-th echo $s_j$'s hidden state feature $h_j$ recurrently after Rectified Linear Unit (ReLU) activation:

$$h_j = ReLU\left(N_m(s_j) + N_h(h_{j-1})\right). \quad (4)$$

This recurrent network attempts to implicitly capture the echo dynamics and fuse features from the preceding echoes. Second, another convolutional layer $N_s(\cdot)$ is used to extract features of the single-echo images $s_{N_E+1}$, $s_{N_E+2}$ and $s_{N_E+3}$, corresponding to the two inversion recovery and the single T2 prepared images, respectively:

$$h_{N_E+1:3} = N_s(s_{N_E+1:3}). \quad (5)$$

Finally, the feature maps of all echoes are updated by fusing $h_j$ and $h_{N_E+1:3}$:

$$h_j = h_j + h_{N_E+1} + h_{N_E+2} + h_{N_E+3} \ (j = 1, 2, ..., N_E),$$

$$h_{N_E+1:3} = h_1 + h_{N_E+1:3}. \quad (6)$$

After the multi-contrast feature fusion, all feature maps are concatenated along the channel dimension and fed into a denoising network based on U-Net (45) to generate denoised multi-contrast images.

**Data acquisition and processing**

All sequences were run on a 3T GE scanner with a 32-channel head coil. Fully sampled k-space data were acquired in 13 healthy subjects following an IRB approved protocol. Voxel size was $0.75 \times 0.75 \times 1 \ mm^3$ with imaging parameters listed in Table S1. Coil compression (46) was applied to the original 32-coil k-space data, generating 8 virtual coils to save GPU memory. A coil sensitivity map was then estimated with ESPIRiT (47) using a centric 20×20×20 self-calibration k-space region. Fully sampled multi-contrast images were computed by taking inverse Fourier transform of multi-coil k-space data and combining them using the obtained coil sensitivity maps to provide labels for network training and result validation. 8/1/4 subjects (2560/320/1280 2D coronal slices) were used as training, validation, and test datasets, respectively. K-space data were also retrospectively sampled on the same test subjects using the learned $R = 8$ under-sampling pattern.

Under-sampled k-space data were acquired in the same 4 test healthy subjects following an IRB approved protocol, using the same imaging parameters as above while using the under-sampling pattern obtained during training on the previous healthy subjects. The same data processing was applied to the prospectively under-sampled k-space data and used as additional test data.

T1 and T2 maps were calculated by dictionary matching. The dictionary containing a 4-time-point transverse magnetization for the mcLARO sequence was generated using a numerical Bloch simulation with sequence parameters in Table S1 (assuming on-resonance condition) and T1 values (ms) in [100:10:2000] and T2 values (ms) in [10:1:200]. QSM was calculated from the mGRE images through fitting the total field map (5,6), removing background field (7), and solving dipole inversion (8). $T2^*$ was calculated using the ARLO algorithm (48).

The training process consisted of two phases. In phase one, weights in the deep ADMM network and sampling pattern optimization module were updated simultaneously by maximizing a channel-wise structural similarity index measure (SSIM) (49). In phase two, the pre-trained deep ADMM network from phase one was fine-tuned with fixed binary sampling patterns. We implemented the training in PyTorch using the Adam optimizer (50) (batch size 1, number of epochs 100 and initial learning rate $10^{-3}$) on a RTX 2080Ti GPU.

**Comparison and statistical analysis**

An ablation study was conducted on the retrospectively under-sampled data to validate the efficacy of the multi-contrast feature fusion (Eq. 6) and the sampling pattern optimization of the three single-echo GRE acqusitions (Eq. 2) in mcLARO. We compared the ADMM reconstruction obtained without either of them (denoted as "mcLARO=00") with those obtained using only sampling pattern optimization (denoted as "mcLARO=01"), only multi-contrast feature fusion (denoted as "mcLARO=10") and both (denoted as "mcLARO=11"). Multi-echo sampling pattern optimization and temporal feature fusion of mGRE images for $T2^*$ and QSM mapping, which had already been validated in LARO (30), were used in the ablation study. When the single-echo sampling pattern optimization was not applied, variable density sampling patterns were designed using a multi-level sampling scheme (51), where sampling pattern of each single-echo was generated independently from a manually designed probabilistic density function. A reference-free image blurriness metric (52) was used to measure reconstruction quality (a score between 0 and 1, with lower indicating less blurring). For each test subject, the first single-echo image from the fully sampled scan was used to segment 114 regions of interest (ROIs) using FreeSurfer (53,54). Bland-Altman analyses (55) were performed to measure the agreement between the regional T1, T2, $T2^*$ and QSM values obtained from the fully sampled and under-sampled reconstructions.

Reference scans for T1, T2, $T2^*$ and QSM mapping were acquired separately using conventional sampling and compared to the mcLARO prospective experiment on the same four test subjects. The sequence parameters of the reference scans are summarized in Table S2. In each subject, 12 ROIs were manually drawn that were contained in the slice coverage of the 2D reference scans, including anterior white matter, caudate nucleus, putamen, globus pallidus, substantia nigra, and red nucleus (one ROI in each hemisphere for each region). Bland-Altman analyses (55) were used to assess the agreement between the regional T1, T2, $T2^*$ and QSM values obtained with the reference scans and the prospectively under-sampled mcLARO scans.

## RESULTS

### Retrospectively under-sampled ablation study

Figure 3 shows an example of the quantitative maps obtained from the fully sampled and $R = 8$ retrospectively under-sampled reconstructions in one representative test subject. In the T1 map comparison (1st row), the moderate noise in the fully sampled T1 map (1st column) was reduced in the under-sampled T1 maps from the mcLARO ablation study (2nd to 4th columns). In the zoomed in T1 maps, the "mcLARO=00" reconstruction (2nd column, without the multi-contrast feature fusion or the sampling pattern optimization modules) showed blurry depictions of the putamen and thalamus. These depictions progressively improved as more modules were added in the "mcLARO=01" (3rd column, with the sampling pattern optimization module), "mcLARO=10" (4th column, with the multi-contrast feature fusion module) and "mcLARO=11" (5th column, with both modules) reconstructions. The blurriness score of the fully sampled, mcLARO=00, mcLARO=01, mcLARO=10 and mcLARO=11 T1 maps were 0.24, 0.32, 0.31, 0.30 and 0.29, respectively, demonstrating improved image sharpness in the ablation study. For T2 and T2* maps (2nd and 3rd rows), the slight noise observed in the fully sampled maps was reduced in the under-sampled reconstructions. No visual differences were observed among under-sampled reconstructions of T2, T2* and QSM in the ablation study.

Figure 4 shows Bland–Altman plots of regional T1, T2, $T2^*$ and QSM values obtained with the fully sampled and under-sampled reconstructions from the four test subjects. For all under-sampled reconstructions, negligible bias and narrow 95% limits of agreement were obtained.

### Prospectively under-sampled reconstruction

Figure 5a shows $R = 8$ prospectively under-sampled mcLARO quantitative maps of one test subject (1st row) and reference scans (2nd row), demonstrating good visual agreement of all the quantitative maps obtained with mcLARO and reference scans. Please note that the visualized anatomy is only approximately similar as head motion may have occurred between acquisitions. Figure 5b shows Bland–Altman plots of regional T1, T2, T2$^*$ and QSM values obtained from the four test subjects, demonstrating small or negligible bias and narrow 95% limits of agreement.

## DISCUSSION

We developed mcLARO as a new learning-based framework for fast whole brain sub-millimeter T1, T2, T2$^*$ and QSM mapping in a single scan. Our ablation study showed the efficacy of the multi-contrast sampling pattern optimization and temporal feature fusion in mcLARO. Our prospective experiment showed comparable quantitative values of mcLARO in the selected ROIs with respect to the reference quantitative scans.

Based the LARO (30) method for mGRE sampling pattern optimization, the k-space sampling pattern in mcLARO, which included both single and multi-echo GRE acquisitions, was optimized independently for each echo and contrast. An optimized spatial incoherency of the learned sampling pattern in each echo was achieved by updating the probabilistic sampling density in Eq. 2 during back-propagation to minimize reconstruction error with respect to the fully sampled images. This was verified by our ablation study in Figures 3 and S1, where incorporating the sampling pattern optimization module improved the sharpness of the putamen and thalamus (4th and 5th columns of T1 maps in Figure 3).

Multi-contrast images acquired by the proposed sequence were naturally aligned and shared similar structural information, which was utilized by the proposed multi-contrast feature fusion module in Figure 2c. Multi-echo feature fusion has been proposed in LARO (30) using a recurrent convolutional module. Based on LARO, multi-contrast image features from all the contrasts were similarly aggregated in mcLARO. Improved putamen and thalamus depiction (3rd and 5th columns of T1 maps in Figure 3) shows the effectiveness of the multi-contrast feature fusion. In addition, noise in the fully sampled images was suppressed in the reconstructed images (zoomed in images in Figure 3). This phenomenon may be explained by the noise2noise (56) experience in deep learning image restoration, where convolutional network is demonstrated to predict an averaged output from a training dataset with unbiased noise.

Prospective results in Figures 5 demonstrate that mcLARO in less than 6 minutes yielded comparable quantitative values to the reference quantitative scans. Compared to other quantitative multi-parametric mapping methods such as MP2RAGEME (1), MR Multitasking (2) and 3D QALAS (40), mcLARO uses a similar GRE-based IR and T2 prepped pulse sequence for contrast variations. The difference is that mcLARO is based on a learning-based framework for both sampling pattern and image reconstruction optimization. Learnable weights in mcLARO are updated to produce better spatial-temporal sampling incoherency, multi-contrast feature aggregation and regularization to improve reconstruction quality under a high under-sampling ratio ($R = 8$). As a result, mcLARO achieves a whole brain sub-millimeter multi-parametric mapping in a shorter scan time compared to other methods, such as MP2RAGEME ($R = 2.89$) (1) and 3D-QALAS ($R = 1.7$) (40). Recently, multi-contrast images in 3D-QALAS were reconstructed using model-based deep learning (29) where acceleration was pushed to $R = 12$. Our method differs in that additional multi-echo GRE is available and a sampling pattern is learned for each contrast. Future work includes pushing mcLARO acceleration to higher under-sampling ratio, such as $R = 12$.

There are some limitations of mcLARO. First, only a limited number of echoes are acquired to capture contrast variation due to magnetization relaxation (Figure 1a). A more suitable way may be time-resolved sampling, such as in MR Multitasking (2) and 3D-EPTI (57), where k-space centers are acquired more frequently to capture the variation. Deep learning has been used for time-resolved imaging with subspace learning (58). Future work may include applying time-resolved acquisition and deep subspace reconstruction to mcLARO. Second, the long scan time of the fully sampled data in mcLARO may introduce motion artifacts, including motion blurring(59). Self-supervised learning via training directly on under-sampled data may replace the possibly motion-corrupted fully sampled labels (60), but direct motion estimation may be needed to reducing blurring(61). Future work includes exploring self-supervised learning strategies in mcLARO. Third, the study subject sample size is very limited. mcLARO prospective study was not applied to patients with pathology not seen in training. The study organ was limited to the brain, and body QSM would require chemical shift correction(62), in addition to motion compensation. Future work also includes testing the generalization ability of mcLARO in organs outside the brain and on patients with new pathology.

# CONCLUSION

In this work, we demonstrated the feasibility of acquiring sub-millimeter T1, T2, T2* and QSM maps in vivo in under 6 minutes. The method uses learned sampling pattern optimization and multi-contrast multi-echo feature fusion in a deep learning approach.

**Figures**

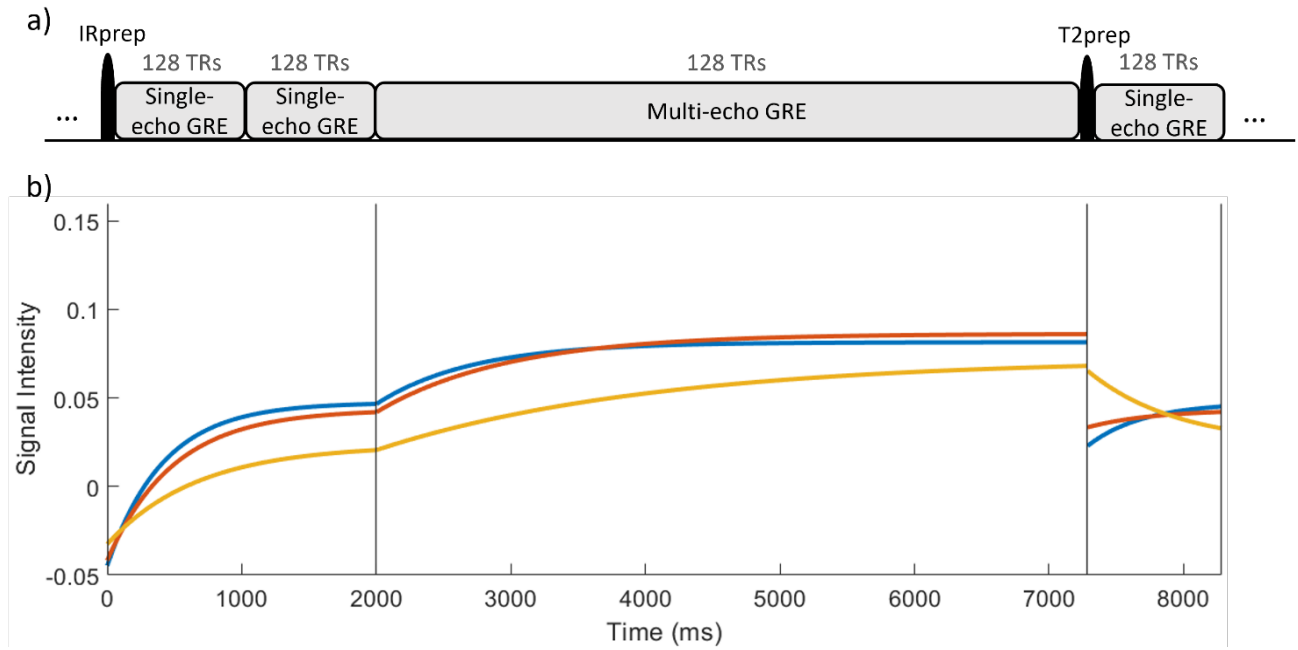

**Figure 1.** a) Schematics of the proposed mcLARO pulse sequence for multi-parametric mapping, which consists of inversion recovery (IR) and T2prep magnetization preparations and single and multi-echo GRE readouts; b) Bloch simulation of the steady state signal of the white matter (T1/T2 = 855/67 ms, blue), gray matter (1264/89 ms, orange), and CSF (T1/T2 = 4000/2000 ms, yellow).

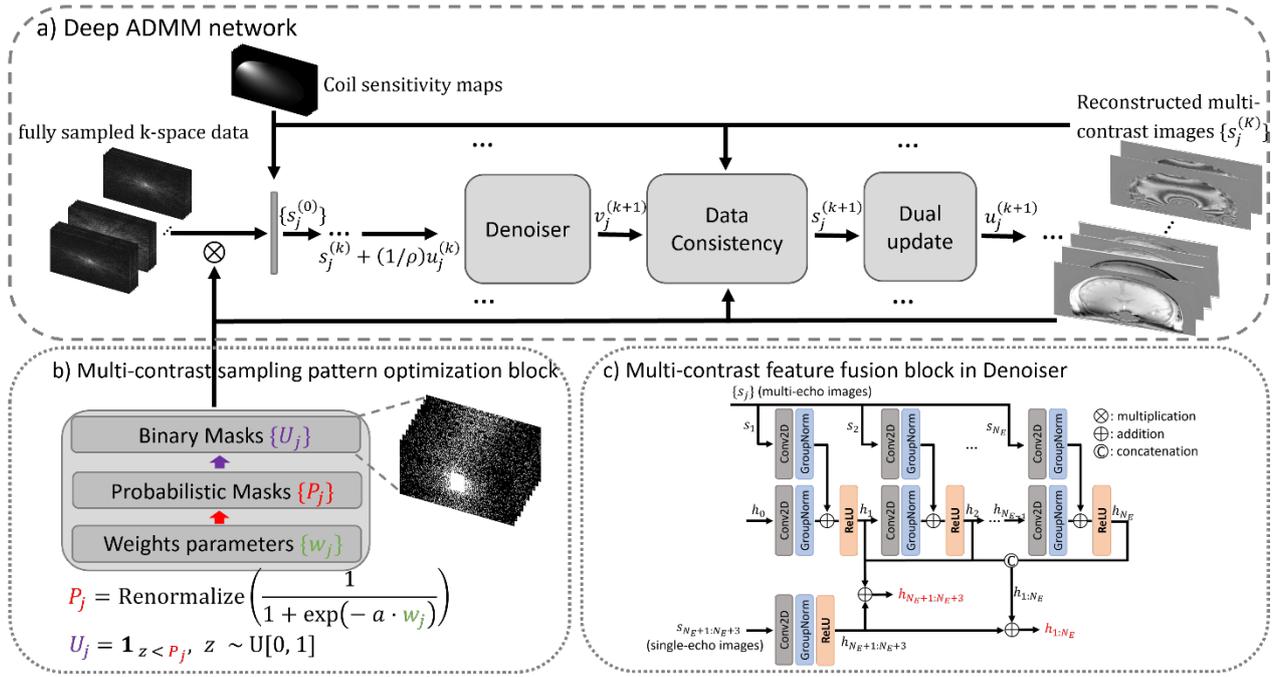

**Figure 2.** a) Deep unrolled ADMM network of mcLARO multi-contrast reconstruction; b) multi-contrast sampling pattern optimization module to learn an optimized pattern from fully sampled data; c) multi-contrast feature fusion module to aggregate information across contrasts during reconstruction.

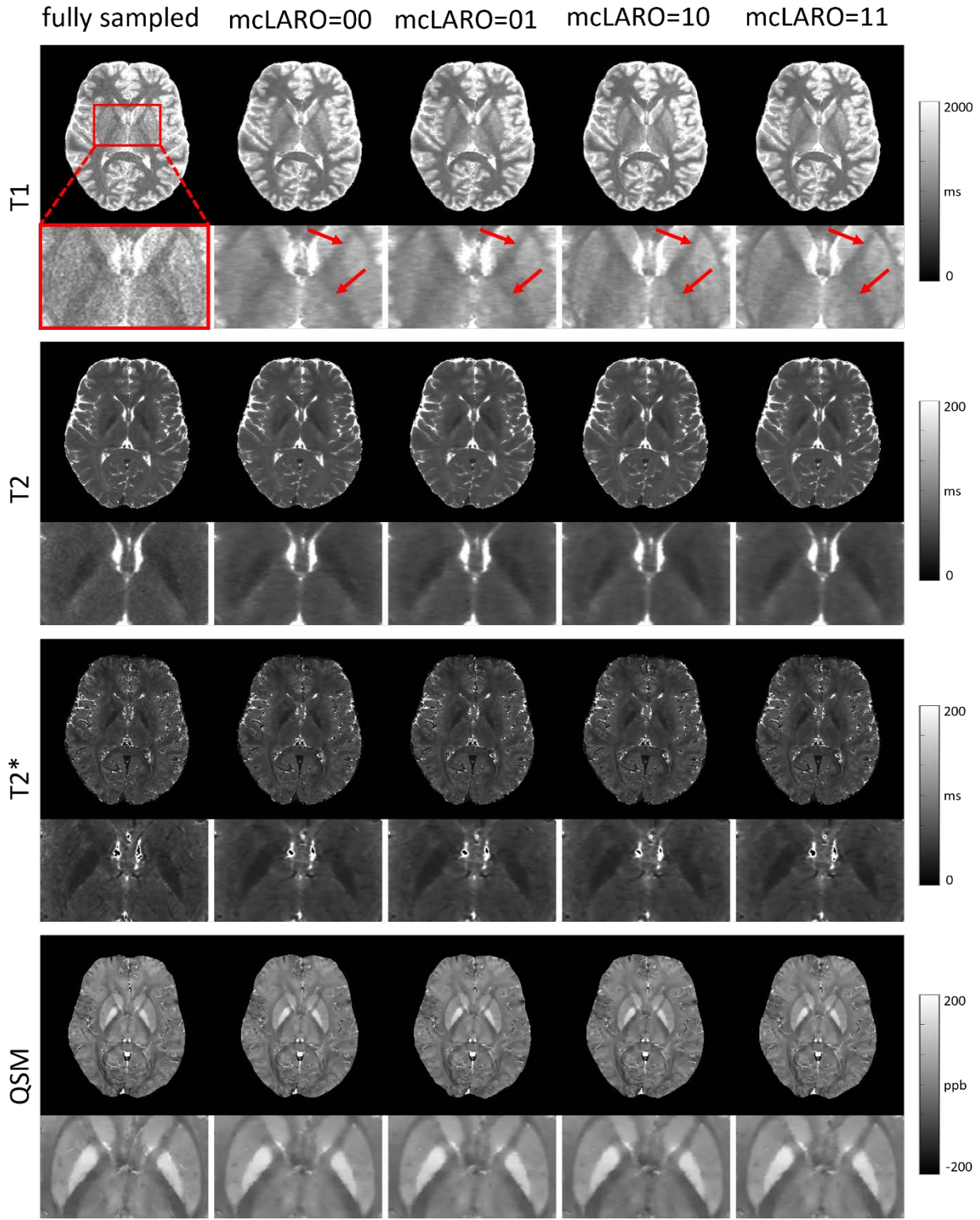

**Figure 3.** Ablation study of the multi-contrast sampling pattern optimization and multi-contrast feature fusion modules in mcLARO with $R = 8$ retrospective under-sampling from the fully sampled data of one representative test subject. For T1 maps (1st row), the noise visible in the fully sampled reconstruction (1st column) was reduced in all under-sampled reconstructions (2nd to 5th columns).

Deep grey matter regions in the zoomed in images (red arrows) were blurry without the two modules (mcLARO=00), but were progressively improved when the sampling pattern optimization (mcLARO=01), feature fusion (mcLARO=10) and combined (mcLARO=11) modules were applied. For T2 and T2* maps, noise in the fully sampled reconstruction was reduced in all under-sampled reconstructions. No visual differences were observed among under-sampled reconstructions of T2, T2* and QSM.

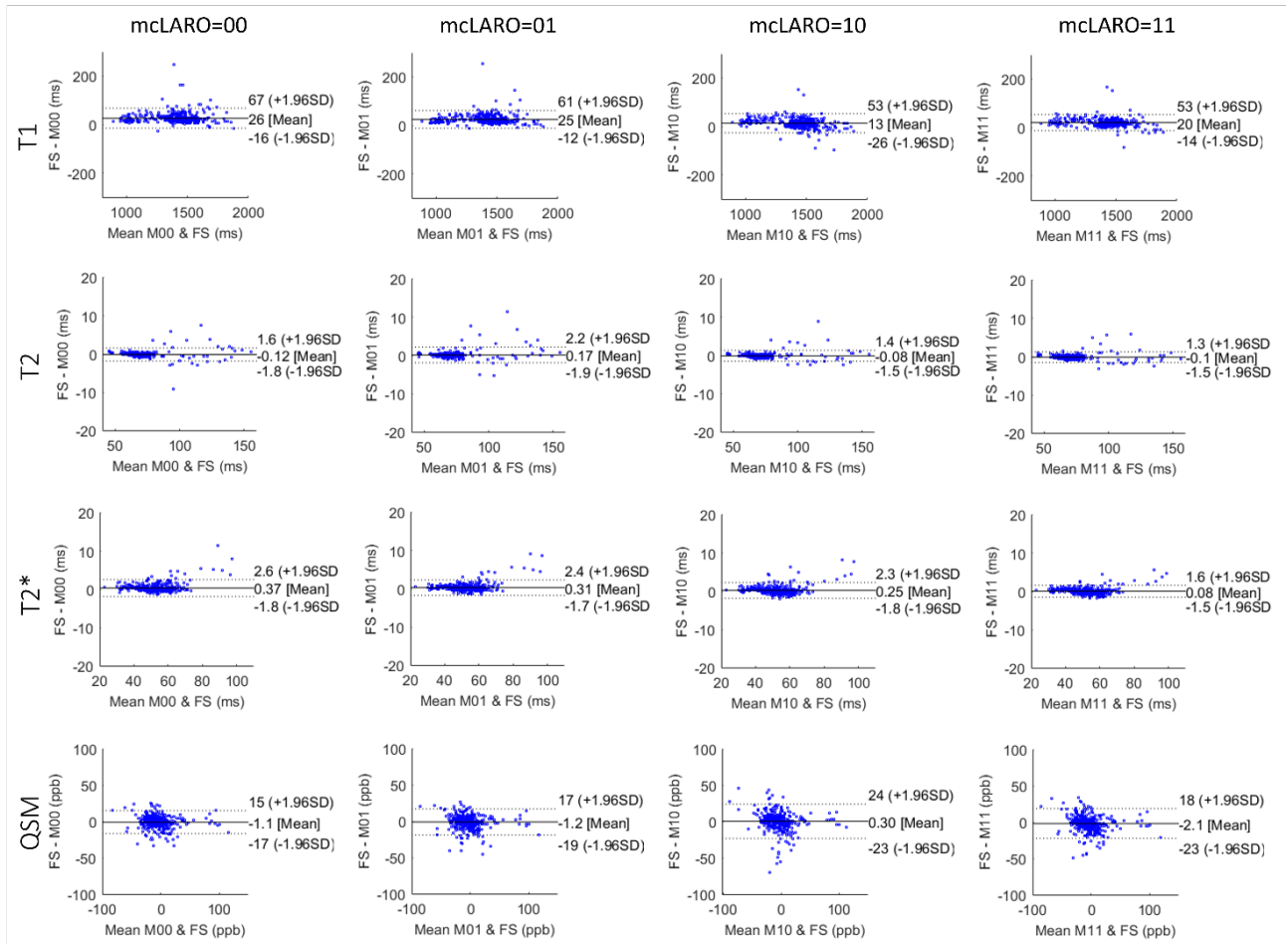

**Figure 4.** Bland–Altman plots of regional T1, T2, T2* and QSM ROI values between fully sampled and retrospectively under-sampled reconstructions on the four test subjects. For all the under-sampled reconstructions, negligible bias and narrow 95% limits of agreement were obtained (FS = Fully Sampled, M = mcLARO).

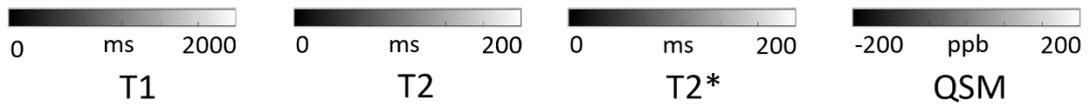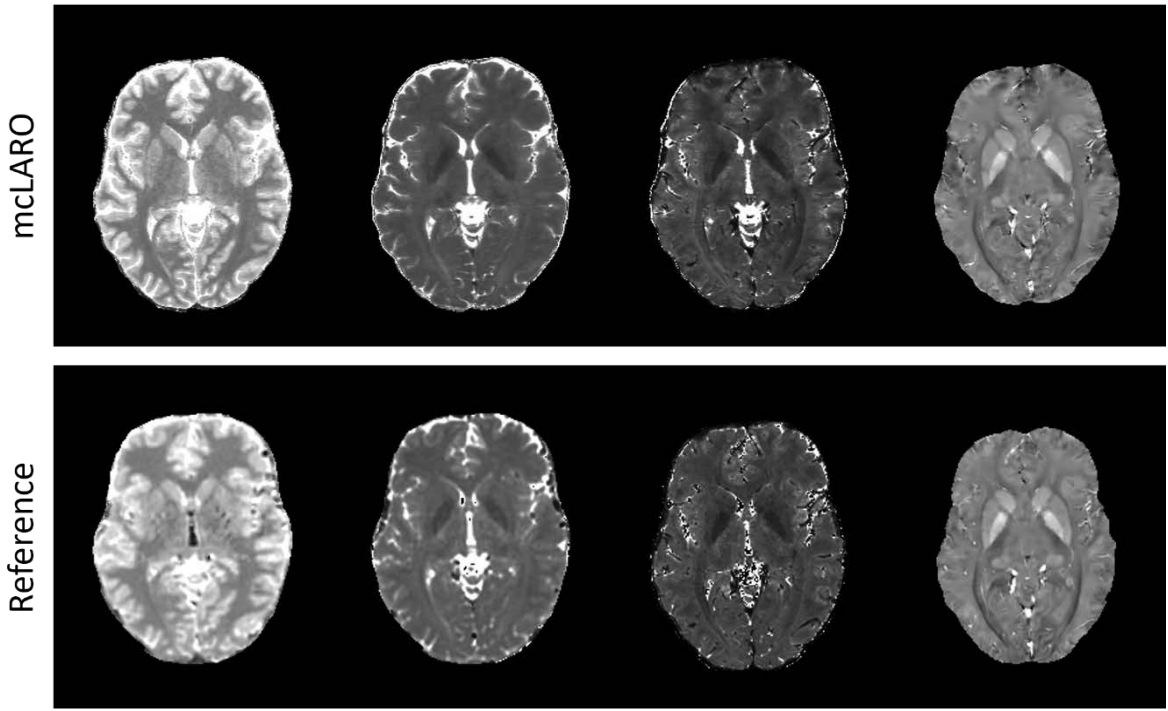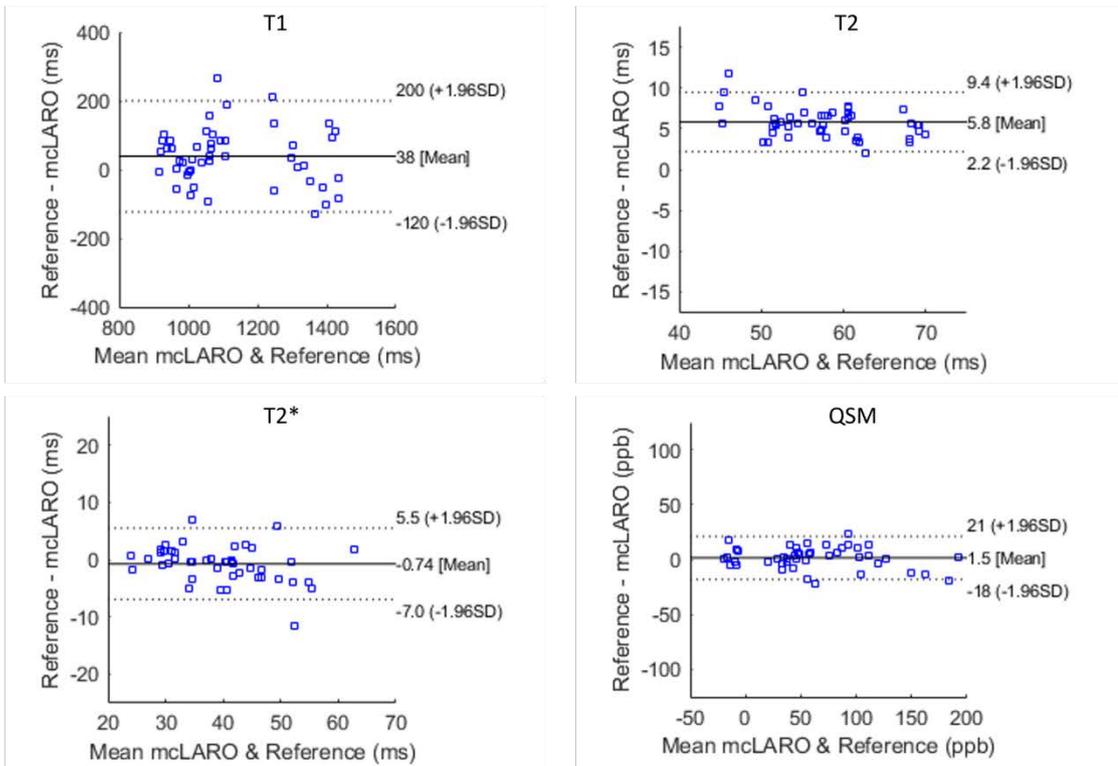

**Figure 5.** a) mcLARO and reference multi-parametric maps. Similar multi-parametric maps were derived from mcLARO compared to the reference. b) Bland–Altman plots of regional T1, T2, T2$^*$ and QSM values obtained with the proposed mcLARO and reference methods from four test subjects. Small or negligible bias and narrow 95% limits of agreement were achieved by mcLARO.

**Supplementary Information**

Table S1. Sequence parameters of mcLARO.

|  | **mcLARO** |
|---|---|
| FOV | $256 \times 206 \times 160$ mm$^3$ |
| Matrix Size | $320 \times 258 \times 160$ |
| Flip Angle | 8° |
| BW | ±50 kHz |
| TR$_{GRE}$ | 7.8 ms |
| TR$_{mGRE}$ | 41.6 ms |
| TE$_{GRE}$ | 2.9 ms |
| TE$_{mGRE}$ | [2.9, 7.7, 12.5, 17.4, 22.2, 27.0, 31.8, 36.7] ms |
| T2prep Echo Time | 85 ms |
| Number of TRs per Segment | 128 |
| Fully Sampled Scan Time | 34 min 30 s |
| R = 8 Under-sampled Scan Time | 5 min 39 s |

Table S2. Sequence parameters of reference scans for T1, T2, T2* and QSM maps.

| | 2D IR-FSE | 2D ME-FSE |
|---|---|---|
| FOV | $256 \times 256$ mm$^2$ | $256 \times 256$ mm$^2$ |
| Matrix Size | $128 \times 128$ | $128 \times 128$ |
| Slice Thickness | 1 mm | 1 mm |
| Slice Spacing | 4 mm | 4 mm |
| # of Slices | 12 | 12 |
| BW | ±15.63 kHz | ±15.63 kHz |
| TR | 3327 ms | 1000 ms |
| TI (TE) | [50, 200, 500, 1000, 2000, 3000] ms | [7.6, 15.2, 22.8, 30.5, 38.1, 45.7, 53.4, 61.0, 68.6, 76.3, 83.9, 91.5, 99.2, 106.8, 114.4, 122.1] ms |
| # of Acquisitions | 1 | 2 |
| Turbo Factor | 8 | ─ |
| Acceleration | 2 | ─ |
| Scan Time | 25 min 42 s | 3 min 28 s |

2D IR-FSE: 2D inversion recovery fast spin echo for T1 mapping; 2D ME-FSE: 2D multi-echo spin echo for T2 mapping

| | 3D mGRE |
|---|---|
| FOV | $256 \times 206 \times 160$ mm$^3$ |
| Matrix Size | $320 \times 258 \times 160$ |
| Flip Angle | 12° |
| BW | ±50 kHz |
| TR | 41.6 ms |
| TE | [2.9, 7.7, 12.5, 17.4, 22.2, 27.0, 31.8, 36.7] ms |
| Acceleration | 2 |
| Scan Time | 11 min 53 s |

3D mGRE: 3D muti-echo GRE for T2* and QSM mapping